# Toward A Normative Theory of Normative Marketing Theory

Ian F. Wilkinson
Discipline of Marketing
The University of Sydney

Louise C. Young
School of Business
University of Western Sydney

## ABSTRACT

We show how different approaches to developing marketing strategies depending on the type of environment a firm faces, where environments are distinguished in terms of their systems properties not their context.  Particular emphasis is given to turbulent environments in which outcomes are not a priori predictable and are not traceable to individual firm actions and we show that, in these conditions, the relevant unit of competitive response and understanding is no longer the individual firm but the network of relations comprising interdependent, interacting firms.  Networks of relations are complex adaptive systems that are more 'intelligent' than the individual firms that comprise them and are capable of comprehending and responding to more complex and turbulent environments. Yet they are co-produced by the patterns of actions and interactions of the firms involved.   The creation and accessing of such distributed intelligence cannot be centrally directed, as this necessarily limits it. Instead managers and firms are involved in a kind of participatory planning and adaptation process through which the network self-organises and adapts. Drawing on research in systems theory, complexity, biology and cognitive science, extensions to the resource-based theory of the firm are proposed that include how resources are linked across relations and network in a dynamic and evolutionary way.  The concept of an extended firm and soft assembled strategies are introduced to describe the nature of the strategy development process. This results in a more theoretically grounded basis for understanding the nature and role of relationship and network strategies in marketing and management.  We finish by considering the research implications of our analysis and the role of agent based models as a means of sensitising and informing management action.

## Introduction

Normative marketing theory involves prescriptions for firm's strategies and planning – implying what firms should do.  There is no normative theory of chemistry physics or biology because chemicals, atoms, plants and animals do not use theories of chemistry, physics and biology to decide what they should do.  But in business this is not so; managers want to understand how and why business and markets work as they do and need theories about how they can develop better marketing strategies and become more effective competitors.  In this article we show how the nature of the strategic problems confronting firms vary depending on the type of environment in which they operate, including the role and importance of the market relations and networks within which a firm is embedded. Firms' strategic planning and behaviour has to learn and adapt to these changing realities and so do our normative marketing theories.

The development of normative marketing theory tends to be portrayed as a development from less to more sophisticated theories.  For example, relationship marketing can be viewed as a development of marketing theory that focused attention on the role and value of continuing relations with customers, as well as other organisations, as a means of gaining competitive advantage, rather than a focus on the marketing mix and individual market transactions (e.g. Anderson et al 1994,

Ford 1980; Gronroos 1994; Hakansson 1982; Sheth and Parvatiyar 2000).  And the subject of relations and networks is now a major area of research and management attention in marketing and other business disciplines.

While the development of normative marketing concepts can be viewed as an evolution of ideas, we argue here that they more fundamentally reflect the adaptation of firms' behaviour and theories of firm behaviour to changing market and competitive conditions.  As part of our analysis, we will show how the nature and role of relations and networks in a firm's marketing strategy varies according to the type of environment in which it operates and creates different types of strategic problems.

We believe it is important for theory and practice to understand the necessary match between normative theories of marketing and the types of environments in which they are effective and appropriate and when they are not. This is because normative theories appropriate for one environment will not work, or have only limited relevance, in other, more complex environments and normative theories appropriate to more complex environments will prove misleading, inefficient and/or ineffective in less complex environments.

The types of differences in organisational environments we focus on here are not of the kind typically found in marketing and business texts.  We are not concerned with changes in the nature and content of specific aspects of the environment, such as the legal framework, overall market demand and supply conditions, or socio-economic conditions; important as they may be for developing and adapting specific strategies.  Instead, we focus on the generic types of strategic problems and uncertainties different types of environments create for firms operating in them; the genotypical characteristics of the environment rather than its phenotypical characteristics (Emery 1977; Glaser 1985).  Specifically, we distinguish between four generic types of environment based on what Emery and Trist (1965) refer to as their causal texture or systemic properties.  This refers to their inherent dynamics and the way different parts of the environment are interconnected and interact with each other over time.  The causal texture of a firm's environment varies by market context and evolves over time.  By this we do not mean that environments necessarily evolve from less to more complex forms over time, though there may be some truth in this.  Many patterns of evolution are possible in different business sectors and a given firm's environment will comprise elements of more than one generic type, giving rise to different types of strategic problems and normative issues.

The paper is organised as follows. First, we describe the four types of environment based on Emery and Trist's (1965) classification.  This is followed by a discussion of the types of strategic problems and normative prescriptions appropriate to each type of environment, including the differing roles relations and networks within and between firms can play.  The nature and problems of planning and adaptation in Type 4 turbulent environments are given special attention because, increasingly, business environments are being described in these terms and because the complexity of this type of environment exceeds the ability of individual firms to cope and challenges traditional normative concepts of firm based strategy.  Next, drawing on recent developments in complexity theory, ecological systems theory and cognition, extensions to a resource or resource advantage-based theory of the firm are proposed relevant to type 4 environments, that incorporate links between resources within and across firms, relations and networks.  This leads to the development of the concepts of the extended firm and soft assembled strategies. Finally, we consider the overall management and research implications arising from our analysis.

## The Types of Environment in which Firms Operate

Emery and Trist's (1965) conceptualization of organizational environments is based on an open systems framework and focuses attention on the types of interconnections that exist between and within a focal system, such as a firm, and the environmental system in which it operates.  The problem of adaptation for a firm is not so much to the ever-varying specific pattern of stimuli encountered but to the underlying structural properties of the environment that produces these

stimuli – for example are they a product of random variation or complex interacting systems? The environment, or extended social field as it is sometimes referred to, is itself conceptualized in terms of its internal dynamics and interconnectedness, its causal texture, not its specific content. Figure 1 shows the basic framework. Four types of interactions or links (L) are depicted, within the focal firm or system ($L_{11}$), within the environment ($L_{22}$), system to environment ($L_{12}$) and environment to system ($L_{21}$). The problem of adaptation is that of matching the properties of the system with that of the relevant environment in which is operates, i.e. of aligning $L_{11}$ with $L_{22}$, which takes place through the processes of learning about environmental changes ($L_{21}$), and active adaptive planning and response ($L_{12}$) (Emery 1999).

**Figure 1 Interactions between a System and its Environment**
(Source: Emery 1977)

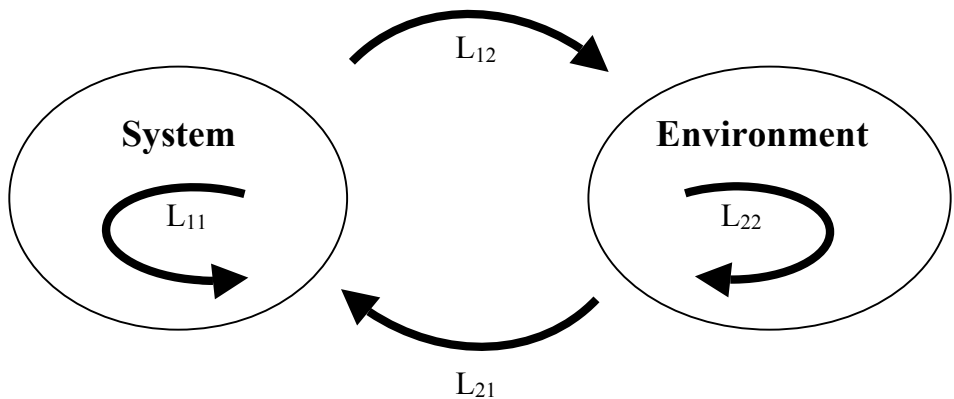

Emery and Trist identify four different types of environments based on their systemic properties or causal texture ($L_{22}$): random placid, placid clustered, disturbed reactive and turbulent. Each type poses different types of problems for planning and adaptation for systems operating in such environments, which may be characterized in terms of one or more of the four types of interactions depicted in Figure 1. The causal texture of an environment is not fixed but may change over time as the relevant social field develops and evolves and as systems and their environment co-evolve. The four types are discussed in order of their increasing complexity but this is not meant to imply that this is some necessary or pre-given stage model of environmental development. First, environments exist within environments, such that a given business context comprises different types of environments with corresponding problems and issues for firms to deal with. Second, the pattern of evolution depends on the co-evolution of firms and environments; on the internal and external interaction processes taking place over time within and among firms and environments (Aldrich 1999).

*Type 1: Random Placid*

In a random placid environment is a market space of opportunities that is randomly scattered and does not change in response to a firm's activities. A firm resembles a small fish in a big pond in which it cannot know what will turn up next, when or from where. In this situation there is no optimal location or market position or strategy to follow as any market position or strategy is equally likely to be beneficial or not and discovering one opportunity or problem gives no guidance as to the nature of other opportunities or problems. Here, no strategy is better than the best tactic and random behaviour, such as tossing a coin to decide the next move in market space, is just as good as any other move. As Emery (1977, p.6) describes it: 'The guiding principle is "catch as catch can" and the characteristic forms of behaviour will necessarily be exploratory and trial and error; vacillating from one hunch to another, even switching to contradictory hunches rather than being persevering or parsimonious.' Learning in such an environment is impossible as any patterns detected are illusions and give no guide for future action.

The focus of adaptation here is the focal firm itself, as all that matters is its resources, reflected in $L_{11}$, and luck (Emery 1977). The environment is similar to what Knight (1921) refers to as risk, or randomness with knowable probabilities, rather than uncertainty, randomness with unknowable probabilities, which is characteristic of other types of environments. Given the random nature of the environment, larger firms or ones that have more resources to sample a larger part of the environment will tend to have payoffs closer to the mean value of opportunities in the environment. This is because they can take advantage of the lower variance among larger samples due to the law of large numbers and the central limit theorem. The mean value of opportunities from larger samples (searches) of the environment will more closely approach the average value in environment as a whole. Smaller search areas (samples) by firms with less resources will tend to have greater variance, so that some will do well by chance and others not. This leads to firm strategies reflected in the principles of massed reserves or pooled risk in marketing, whereby central stocks and reserves held against uncertain demands gain cost efficiencies, such as in insurance (Dixon and Wilkinson 1986).

Normative theory is limited in such conditions. There can be no experts with superior knowledge and the only advice that can be offered is that firms should not waste time and effort trying to improve their strategies because any strategy is as good as any other and only size counts in absorbing the effects of such randomness. Some financial markets may appear to function in this way and random walk investment strategies have proved to be nearly as good as more sophisticated ones in some cases.

Note that random placid is not the same as a perfectly competitive market. In such a market consumers and suppliers are homogenous and perfectly informed. But customers are randomly paired with suppliers as there are no transaction costs to reduce or transaction benefits to gain from repeated transactions. Hence there is no value in forming relationships between customers and suppliers. Firms are price takers and can sell as much as they want at the prevailing price without affecting the price. The optimal strategy is to produce the demanded good and offer as much as maximizes profit i.e. when marginal cost equals the price and U shaped cost curves are assumed which limit a firm's size.

*Type 2: Clustered Placid Environments*

In clustered placid environments opportunities are not random but the environment is otherwise similar to the first type. The environment does not change in response to a firm's actions and it is neutral in the sense that it is not deliberately designed to prevent a firm's survival or to actively seek to harm it. The existence of non-random aspects to the environment makes learning possible and valuable. Adaptation focuses on the firm's ability to understand and respond in appropriate ways to its environment, i.e. on $L_{11}$ and $L_{12}$, through the identification of more and less preferred market positions and strategic moves. 'In an environment where things are clustered together in time and place it is possible for some things to act as cues for the co-occurrence or subsequent appearance of other things and for others to be seen as co-producers of certain desired effects' (Emery 1977 p.7). The small fish in the big pond can now learn about its environment and move to preferred locations. In marketing terms, a consumer orientated market strategy is called for in which firms seek to understand and respond to consumers' needs through market segmentation and positioning strategies. Sources of supply are also not randomly scattered and require similar differentiation and positioning strategies to gain access to those best suited to target customer segments. We are able to rely on market research responses as clues to the actual behavior of consumers (and supply sources), which contrasts with a random placid environment where there would be no relation between research results and actual behaviour. In addition, because the environment is placid, some forms of cooperation among complementary market actors with non competing goals is possible, such as working with customers, suppliers or distributors to improve products and services. However, there are still potential conflicts with customers in terms of what the best use of the firm's resources are.

Such an environment corresponds to a situation in which firms can choose their strategies so as to optimise returns based on an evaluation of the relative value and likely outcome of different alternatives. Because firms operate in an environment that is potentially knowable, the resource based theory of the firm (Barney 1991) explains why firms succeed or fail in these environments; the firm is limited only by its own resources and competences in researching and responding to market opportunities. In principle, optimal strategies can be devised using various types of research, analytical and optimizing techniques, which take into account the costs and benefits of alternatives and much of neoclassical economics has evolved to solve these types of constrained optimization problems (Denzau and North 1994).

As Lane and Maxfield (1996) point out, the methods of solving this type of strategic problem have been and still are the dominant modes of business training and strategic planning. In marketing these techniques are well represented in terms of marketing engineering (Lilien et al 2002), which includes methods to segment markets, to identify preferable market positions and to design profitable market offers (Lilien and Rangaswamy 1998; 1999). As Lilien et al (2002) argue, marketing engineering links the power of computers with data, knowledge and judgment to assist managers' decision making. Managers combine use of these techniques with their ability to identify acceptable trade offs among conflicting goals and the quality of their assumptions about aspects of the environment where research is too costly, limited or biased.

*Type 3 Disturbed Reactive Environments*

Disturbed reactive environments are clustered environments that are not placid. There are two or more competitors or firms of the same type, such that they all cannot simultaneously achieve their goals. It is a zero sum game. The fish is no longer small or alone. A firm's actions affect its competitors directly or indirectly through its interactions with the environment (e.g. meeting needs of customers so they are not available to competitors, competing for sources of supply). The environment is not placid as there are competitors who can gain at other's expense and, therefore, will directly or indirectly try to harm it. Moves by a firm to a preferred location in market space can be disturbed by the reactions of other firms, as when two firms identify the same market niche as an opportunity. In this type of environment adaptation focuses on $L_{11}$, $L_{12}$ and $L_{21}$. $L_{21}$ becomes a focus of attention because of the need to respond to the intentions and capabilities of competitors.

The relevant theory of competitive advantage in disturbed reactive environments is that of resource advantage theory (e.g. Hunt and Derozier 2004; Hunt and Morgan 1997), where resources are defined broadly to include any means by which valued products and services can be created and delivered. This is an extension of the resource based theory of the firm and reflects the need for firms to offer products and services to customers that are preferred to competitors, i.e. are perceived to offer more value. A competitive as well as a consumer orientation is required. Thus it is not the existence and control of resources per se that leads to firms being able to serve customers better but the extent to which they result in differential advantages. The importance of different resources for generating competitive advantage depends on how valuable, rare, inimitable and non-substitutable they are (Barney 1991).

Resource advantage theory is a normative theory that provides a framework for guiding firms in developing competitive advantage and for diagnosing sources of competitive disadvantage. To determine its strategy a firm has to understand the structure of the market, which is assumed to be stable and unaffected by competing firms' actions, including the resources that are important for creating and delivering value, the resources other firms have, and how they are likely to behave and respond.

The game of business here resembles a game of chess or war. At any stage in the game a player can assess their resource advantage in terms of the types of pieces they have compared to their opponent and the positions they occupy relative to opponents. This affects the types of moves that can be made. But even though the rules as well as the power and behaviour of individual pieces is fixed, there is no one best strategy; it depends on what other players' resources and strategies are. Competition, like chess and war, is a process of interaction that takes place over time. At the same time player A is acting and responding to the resources and actions of player B, player B is doing the same concerning player A. Only if each perfectly predicts the behaviour of the other will expected outcomes occur. This only tends to occur in special circumstances, when the alternatives available are severely constrained, as in end games in chess. Usually, competitors have many possible alternatives to consider and in these situations, as Brandenburger (1998) has pointed out, even in a simple two person game the complexity is fundamentally infinite. Each player is responding to what they think the other will do and think and what they think the other thinks they will do and think and so on in an infinite regress.

There is another dimension to competitive interaction in disturbed reactive environments. A new type of strategy emerges that involves out-maneuvering a competitor in order to achieve a strategic end, what Emery (1977) calls 'operations'. But competitors also can engage in misleading and deceptive conduct to confuse others about their actual strategies and intentions. In other words competitor opportunism exists, which Williamson (1975) defines as self-seeking behaviour with guile.

As each firm is anticipating and responding to others, that are in turn anticipating and responding to them, complex patterns of behavior emerge that are not predictable in advance, resulting in unavoidable uncertainty in the Knightian sense. No firm is in control of how competition and outcomes develop. Outcomes are co-produced over time through the pattern of interaction among players' strategies, just as in a game of chess. There are no optimal strategies in such circumstances as they are contingent on the strategies of others. Instead, 'strategic options need to be formulated more in terms of power to meet competitive challenge than simply in terms of achieving optimal location.' (Emery 1977 p.9). The power to meet competitive challenge depends on the ability of firms to respond to different circumstances and Alderson (1957, p.51) identified a key normative principle relevant for guiding interactions in such conditions - the Power Principle. 'An individual or organisation, in order to prevail in the struggle for survival, must act in such a way as to promote the power to act.' In other words a firm tries to take into account how current decisions enable or constrain future decisions and tries to avoid committing resources in such a way that unduly narrows future options and flexibility. This principle is analogous to the type of dominant strategy that exists for all sophisticated competitive games such as Chess or Go, which involve getting your opponent to commit themselves while remaining still flexible yourself (Padgett and Ansell 1993).

Marketing engineering and research techniques are still useful in this environment but they must be augmented by judgements of competitor and environmental responses as well as the focal firm's own future responses. Here the quality of management judgments become relevant and softer forms of data are likely to be important, such as rumour, espionage and gossip.

Competitive theory in Type 3 environments is the realm of Game Theory in which the actions of each player are modeled over time under different assumptions in order to identify better performing strategies. Under various assumptions and game conditions, superior strategies may be identified, including Nash equilibria, in which each player is willing to persist with their strategy so long as others do. But such equilibria are not necessarily optimal or stable. Simulation experiments based on repeated prisoner dilemma games, which involve a mix of competition and cooperation, show how different outcomes can emerge depending on the mix of strategies being played and the starting conditions (Dixit and Nalebuff, 1991). Evolutionary models that allow firms to learn and adapt their strategies over time based on the outcomes of earlier interactions show some interesting results, including the potential emergence of cooperative behaviour among groups of competitors (e.g. Axelrod; 1984, 1997; Lindgren 1997), although others are more skeptical (e.g. Binmore 1998). Such models are far from providing clear directions for management action except in particular types of market auctions, where behaviour is constrained by known rules.

Relationship marketing concepts can play an important role in Type 3 environments. Relations with customers, as well as other types of organisations are potentially important resources that firms can use to create and sustain differential advantage by reducing transaction and coordinating costs, designing and adapting products and services, protecting themselves against competition and gaining valuable information about competitors and the market environment. (e.g. Dyer and Singh 1997; Gulati, Nohria and Zaheer 2000; Kanter 1997). Relationships as resources are particularly valuable because they are hard to duplicate or buy by competitors (Dyer and Singh 1997; Andersen et al 2001). Developing relations with competitors are also potentially valuable in order to open up, control or suppress market competition and to develop acceptable market and industry standards (e.g. Grunstrom and Wilkinson 2005; Welch et al 1997) `

*Type 4 Turbulent Environments*

Turbulent environments are dynamic environments like disturbed reactive environments but the dynamics arises not only from the actions and reactions of competitors but from within the environment itself. The actions of firms not only affect competitors, they also have direct and indirect impacts on other aspects of the environment. The environment comprises many types of purposeful, interconnected, interdependent, interacting systems and evolves in response to the actions and interactions of those involved, leading to the emergence of new types of actors, relations and networks.

The game of business in turbulent environments is like a game of chess with many players who can compete as well as cooperate; a game with no fixed rules, in which the pieces have goals of their own and interaction among pieces and players can change over time as a result of previous interactions, new types of pieces can arise from recombinations of existing pieces and through chance mutations, and where there is a no set goal by which winners and losers are clearly defined. Today's winners may be tomorrow's losers and vice versa and the game never ends. Over time there is a continuous co-evolution of players and patterns of interaction, where each player constitutes part of each other's environment, in which the fittest survive in the context of the environment that is co-created by the fittest, or as Stuart Kauffman (1996) describes it: 'The winning games are the games the winners play' (p79).

A turbulent environment is a complex adaptive system (CAS). Such systems arise in biochemical, biological, ecological, social systems as well as economic and business systems (Arthur et al 1997). The behaviour of a CAS depends on the way the parts are interconnected, not just the characteristics of the individual parts. No participant is in control. Instead, order and large scale structures, emerge in a bottom up self-organizing way from the micro interactions taking place among individual actors. Interactions and interdependencies among the parts of the overall system make its behaviour highly non-linear and impossible to predict. Outcomes are very sensitive to starting conditions and, because the outcomes of any actor's behaviour depend directly and indirectly on the behaviour and responses of many others, outcomes cannot be traced back to the acts of individual actors. 'An organization reacts to the actions of others that are reacting to it. Much of what happens is attributable to those interactions and thus is not easily explicable as the consequence of autonomous action' (March, 1996 p.283).

Managers' intuitions regarding how to effectively participate in such non-linear systems may not serve them well, as traditional approaches to solving managers' problems are based on mathematical methods and approaches designed to deal with linear systems that can be decomposed into independent subparts that can be studied and understood in isolation and for which analytical solutions are possible. But as Robert May (1976, p.467), a pioneer of non-linear systems analysis points out: 'The mathematical intuition so developed, ill equips the student to confront the bizarre behaviour exhibited by the simplest of discrete nonlinear systems… Yet such nonlinear systems are surely the rule, not the exception, outside the physical sciences.'

In some ways this environment resembles the random placid environment and may appear to be so for an individual actor. But it is not random. It obeys rules of behaviour and interaction that defy closed form solutions, even if we could write down the underlying behavioral equations. It is an example of strong uncertainty in the Knightian sense (Denzau and North 1994). The focus of adaptation now includes all types of firm and emvironment interactions i.e. $L_{11}$, $L_{12,}$ $L_{21}$ and $L_{22}$ : 'In turbulent environments adaptation is not possible unless somehow one comes to grips with the $L_{22}$' (Emery 1977 p 10).

How does a firm come to grips with such a complex and dynamic environment and what kinds of normative theories are relevant to guide the behaviour of participants in such systems? As we will argue in the following section, one way this is accomplished is not through the planning and strategic actions of individual firms but through the sensing and response of networks of firms co-created through their actions, interactions and relationships over time. Such networks are more able to recognize and cope with the uncertainties involved and call for different types of competences and skills on the part of participant firms in order to contribute productively to and benefit from the self-organising process by which such networks are created and changed.

## Normative Marketing and Management Theory in Type 4 Environment

### Turbulence and Type 4 Environments

The term turbulence is often used to describe environments in which individual firms face strong uncertainties because of the complex nature of environments resulting in rapid and unexpected change. Indeed, Mintzberg (1993) points out that turbulence of this kind has tended to be ascribed to the current environment in every age and that the patterns and directions of change only become clearer in hindsight. Here we wish to make a distinction between environments in which individual firms may face significant uncertainties and environments that are inherently unpredictable because of the degree of interdependence and interactions taking place in the overall socio-economic system of which individual firms are but one part.

Whether a particular environment is a Type 4 is not determined simply by the degree of uncertainty individual firms face, which can arise for various reasons (e.g. Lane et al 1996). For example, hypercompetition (D'Aveni 1994, 1999) is a source of extreme uncertainty that results from a rapidly increased pace of change in forms and types of competition and is often confused with turbulence. Firms deliberately try to disrupt markets and undermine rival's sources of competitive advantage in an ever escalating race to achieve temporary rather than sustainable forms of competitive advantage. Temporary partnerships with other organisations can play a part in this process of seeking out and exploiting ways of disrupting competition for short-term advantages. But as Selvey et al (2003) point out this is an extreme form of a disturbed reactive environment, not a Type 4 environment. The game of business is still a continuous succession of prisoner dilemma type games played among rivals and the focus is on ways to compete, not on any indirect consequences for other parts of the system, such as the near breakdown of the heath care system in some parts of the US or the waste and polluting effects of industry (Selvey et al 2003; Diamond 2004).

Effective strategy in Type 4 environments requires that second and third order effects of actions are considered because they have significant direct and indirect feedback effects on the competitors and the larger system in which they operate. However, the ability to recognize, understand and cope with these effects is beyond that of the individual organization. It is also beyond the abilities of centrally directed and coordinated networks of relationships, such as strategic networks (Gulati, Nohria and Zaheer 2000; Jarillo 1988) or network forms of internal organisation (e.g. Achrol and Kotler 1999; Ghoshal and Bartlett 1990). Although a centrally directed network permits the lead actor to avail themselves of various types of collaborative advantages via the resources of others it can access and help co-create (Kanter 1994), the overall variety of responses is necessarily restricted by the vision and schemas of the lead actor. Network forms within an organization suffer even more from this problem, as the network is constituted within the boundary of one firm.

Strategic networks and network forms of organisation are essentially types of response to disturbed reactive environments (Type 3). They focus on rivalry between networks as opposed to individual firms. But Type 4 environments refer to conditions in which the environment is beyond the capacity of an individual firm to deal with, no matter how internally organized it may be and no matter how well it is able to orchestrate the behaviour and interactions of other organisations in its strategic network. In any period there will always be winners and losers but no one firm or strategic network is able to cope with all possible contingencies through time.

Type 4 environments are typified by periods of radical restructuring of industry and firm boundaries and discontinuous change, such as were wrought by the coming of the railways, telephone, electricity and the internet. In these cases technological developments disrupted previously established forms of interaction, competition and industry structure until new forms emerge to take their place. An illustration of this type of change is provided by Lane and Maxfield (1996) in terms of the evolution of the firm ROLM and PBX computer based systems. They show how the impact of new technologies led to the emergence of new types of products, actors and relationships that could not be foreseen, planned or intended. Existing customer-supplier relations were undermined, new patterns of interaction and relations formed within and between firms in the industry. The knowledge and learning that co-developed between firms as a result of these interactions led to radical and unexpected adaptations of strategy including redirections in terms of both ends and means.

We should re-emphasise that environments with different types of causal textures can co-exist because there are environments within environments (Emery 1998). Hence for some sectors and regions of industry and for some dimensions of a firm's strategic action space, non Type 4 environments may exist such that they can follow normative strategies borne of these 'simpler' environments or sub-environments. For example, in the PBX example referred to above competitive games were still played out between particular actors over time (Type 3 environment), customer preferences varied over time and place calling for variations in systems, products and solutions adapted to particular requirements (Type 2 environment), and chance events dictated why particular actors or relations gained or lost at particular times and places (Type 1 environment).

But how does a firm cope with a type 4 turbulent environment, which is inherently unpredictable a priori and in which opportunities and outcomes emerge in a self organizing way from the local patterns of interaction, learning and adaptation taking place? Should they continue to try to construct, reconstruct and direct internal and external networks in order to seek at least temporary competitive advantage? Or are new types of strategies called for? A key insight is that *the relevant unit of coping and response is not the individual firm or lead actor but the network itself.* Hence a firm has to learn not only how to adapt its interactions with the environment and competitors but also how to effectively participate in and learn from the networks of relations in which it is actually or potentially embedded.

A useful analogy is the way social insect colonies behave (Bonabeau, Dorigo and Theraulaz 1999). In social insect colonies there are no leaders. Instead, coordinated and directed action takes place through 'rules' of direct and indirect (e.g. via pheromones) interaction with other members of the colony and the environment. These rules have evolved in the context of their particular environment in order for the colony to survive and are hardwired in to their genes. Directly and indirectly their individual actions and responses co-construct the ant or bee colony, including the way it houses itself, breeds, takes care of its young, searches for and shares food, and how it recognizes and responds to opportunities and threats inside and outside the colony. All this is done without any central direction (Beekman and Wilkinson 2004).

If the environment changes so that existing rules of interaction do not produce viable colony behaviour, a struggle for survival will take place. Colonies will take different evolutionary paths in their 'search' for a new, viable set of rules of interaction, which are affected by genetic variations among colonies which give them different starting conditions, by historical circumstances and by path dependencies and random mutations. Eventually, existing species become extinct or take on new forms and new types of species emerge that may redefine what we mean by ants and bees. We and the ants and bees do not know how many evolutionary stable interaction strategies may exist and their number may well be quite limited, as studies of evolutionary convergence suggest (Morris 2003).

Do we have any normative theories that would help an individual ant or bee cope with these conditions?  No, each ant does not have the resources or knowledge to cope alone; their behavior and its outcomes for them and the colony are unknown and unknowable for the individual ant, as they depend on what other ants do and the conditions they face.  Moreover, even for a fully informed outside observer, the outcomes of collective ant behaviour and interactions are not knowable because the colony is a complex adaptive system; a non-linear system in which many possible equilibria exist to which the behaviour of the colony is attracted (hence the name attractors) depending on starting conditions, the type of stimuli experienced and random elements (e.g. Bonabeau,  Dorigo and Theraulaz 1999; Omerod 2001).

*Strategies for Type 4 Turbulence*

While an ant or bee cannot alter its own rules of interaction, humans and firms actively seek to understand and manipulate their interactions with the environment and each other.  They can change both their interpretation of themselves and their environment and - at least to some extent – the way they interact with others.  Such adaptations to turbulent environments can be both maladaptive or adaptive.  First we consider maladaptive responses.

*a) Maladaptive Responses and Type 5 Environments.*  Coping with turbulence involves attempts to adapt to or reduce the level of relevant uncertainty.  Emery (1977) identifies three types of responses to turbulence that are maladaptive: superficiality, segmentation and dissociation.

*Superficiality* involves reducing the relevant uncertainty by lowering emotional investments in the ends being pursued.  This is reflected in managers following ritualistic formal planning procedures they do not believe in but do not know what else to do (Stacey 1996). For example, Mintzberg (1993) points out how formal planning can become a substitute for control in large organisations rather than a means of control; an illusion that reduces management anxiety, allowing them to sleep better at night. Planning becomes an end in itself and a form of public relations that avoids dealing with the essential uncertainties confronting the firm.

*Segmentation* involves breaking the system into smaller, less complex systems that are insulated from each other.  Emery (1977, p.31) sees this in terms of  'the enhancement of ingroup-outgroup prejudices as people seek to simplify their choices.'  In business, industry protection schemes, trade barriers and trade blocks are attempts to reduce interdependencies and protect local business communities.  The recent strong protests at world trade forums also reflect people's beliefs that reducing the international interactions among economies and firms can reduce the complexities and uncertainties for individual nations and firms.  Loyalty schemes, cooperatives and tied relationships can also be manifestations of segmentation.

Lastly, *dissociation* is the denial of the potential contribution of coordinating with others over individual selfish action and is 'characterized by indifference, callousness and cynicism toward others and to existing institutional arrangements' (Emery 1977, p.32).  In business the 'greed is good' type slogan represents this as well as a strategic focus on business relations with an adversarial and zero-sum perspective in which the problem is primarily to reduce dependence, gain power, guard against opportunism and supress conflict.  As one manager put it: 'The conventional wisdom is that business is war, cooperation is for wimps, and winning is everything.  Boundaries must therefore be firmly established and defended, communication restricted.' (James 1994)

These maladaptive responses have led some to propose a Type 5 environment or hyper-turbulence:  'the condition in which environmental demands finally exceed the collective adaptive capacities of members sharing an environment' (McCall and Selvey 1984, p. 460).  Such an environment results from participants' attempts to reduce or avoid the inherent uncertainties through one or more of Emery's maladaptive strategies.  Such strategies would dampen the turbulence and uncertainty by reducing the degree of interconnectivity in the environment. And research on the stability and behavior of complex systems shows how changes in the degree of connectivity has important effects, moving a system from regimes of chaotic to more ordered behavior (Gardner and Ashby 1970; Kauffman 1992; May 1972, 1974). The maladaptive strategies reduce interdependence by partitioning the environment into separate subsystems with very restricted interaction, what McCall and Selvey (1984) refer to as social enclaves and social vortices. Those that are able to collectively cope with the sub-environment, as defined by their own existence and separation, form social enclaves. Social vortices are comprised of those who cannot cope, out-groups that are relegated to the margins of society. The examples given are South Africa before the end of Apartheid and the decoupling of segments of US and other societies into ghettos, and survivalist communities.

*b) Adaptive Responses to Type 4 Environments.* The challenge posed by turbulent environments has led to various proposals regarding appropriate ways for managers and firms to respond. These focus on the importance of participating, learning and knowledge development (Senge 1990), adaptive strategies (Wilkinson and Young 2002) and improvisation (Cherlaru et al 2002; Moorman and Miner 1998). Such strategies recognize that firms are fallible; they have limited information and operate in a complex adaptive system with many possible types of outcomes that are a priori unknown and unknowable. Therefore a chosen strategy is a kind of hypothesis to be changed, both in terms of targeted outcomes and means, in response to feedback as the firm muddles through (Lindblom 1959). In this process of muddling through the firm changes its plans in a continuous way as new ends and new opportunities as well as new means to reach existing ends are discovered, recognized and responded to (Collander, 2003). The patterns of interactions and relations managers and firms have, directly and indirectly, with others in and outside the firm play a key role in this process as they are a source of learning, adaptation and response.

The role of relations is exemplified in Lane and Maxfield's (1996) concept of generative relationships. These shape the way a manager or firm makes sense of its environment, including the relevance of different types of actors, their actual and potential roles and functions, and the way it acts in relations to others (Lane et al 1996). In the PBX case the example is given of the impact of the interactions between technical managers (TM) in customer firms and ROLM, a supplier of new types of computer based systems. These interactions affected the way ROLM understood and approached the market in terms of target market segments, products and systems, it altered the position of TM managers in their firms, it affected the interactions among TMs leading to new customers for ROLM and undermined relations between TMs and other suppliers. The problem is that such generative relations and their outcomes cannot be anticipated in advance or predicted from a knowledge of participants; they emerge through participating in such relations over time (Lane et al 1996). Hence the firm has an ongoing task of monitoring, interpreting and reinterpreting existing and potential patterns of interactions and their meaning.

Emery (1977) also focuses on the key role of relations and networks as the means of coping with turbulent environments. He explains this in terms of the emergence of widely shared values and ideals between organisations and how these relate to the fundamental dimensions of choice. These values and ideals are not some new age version of business philosophy and they are not achieved in the same way as other types of business goals such as market share or profits. Instead they act as guides to the choice and adaptation of goals and objectives in the face of the fundamental uncertainties of turbulent environments; they are kinds of shared mental models that help in the choice and adaptation of our goals (Danzau and North 1994).

> 'Ideals enable people:
> 1.  To maintain continuity of direction and social cohesiveness by choosing another objective when one is achieved or the effort to achieve it has failed;
> 2.  To sacrifice objectives in a manner consistent with the maintenance of direction and cohesion.' (Emery, 1977, p69)

The four dimensions of choice used by Emery as the basis for identifying appropriate values and ideals are probability of choice, probable effectiveness, probable outcome and relative value (Ackoff and Emery 1972).

*Homony and probability or scope of choice and response.* The first ideal relates to improving the probability or range of choice by increasing *homony*. At a personal level this involves more closely relating to others such as neighbours, workmates, etc., through which people will improve their range of choice. 'The experience of others is a prime source of 'familiarity with' the world, and 'the others' are usually best able to provide access to a wider range of course of action.' (Emery 1977, p72). In a similar way, improved coordination and cooperation between and within firms is a means for expanding the action possibilities for all.

*Nurturance and Probable Effectiveness.* The second ideal focuses on resource creation and development through relations and networks. Specifically it focuses on promoting the ability to act and the effectiveness of our actions and responses through cultivating and developing our own as well as other's competences and resources. We are concerned with growing the garden (productive relations and networks) as opposed to just picking the flowers (using relations to meet current objectives). This can be viewed as an extension of Alderson's Power Principle to involve the role of others' actions and capabilities in promoting our own.

*Humanity and Probability of Outcome.* The probability of outcome derives from the first two components of choice and thus presumes homony and nurturance. Here the focus is on the use of others as ends versus means or as Norbet Weiner expressed it 'Towards the human use of human beings' (Emery 1977 p75). At any time firms and people have many different objectives they may pursue. Choosing among them according to humanitarian ideals means not necessarily choosing the best in terms of achieving economic or technical goals but one that fits our nature. Emery sees the ultimate guide here in terms of 'good for whom' and interprets this in terms of how individuals are affected not organisations. Organisations are merely the local environments or habitats within which individuals seek out their ideals, organisations themselves are not ideal seeking.

*Beauty and Relative Value.* The last ideal seems at first remote from the business of business but it is not; it deals with the interrelations of purposes or ends themselves. Ideals only emerge in social systems in which inherent uncertainties arise from the interlocking purposes and goals of the actors involved; from the interdependencies that exist among them. Beauty is used as the term for a higher order rationality, which somehow rationalizes and synthesizes the inherent conflicts of interlocking purposes within a higher rationality or order of ends. In regard to business systems the higher order logic would seem to refer to the operations and maintenance of the overall market directed process itself and to the protection and maintenance of the material, biological and social environment. It is through the market process, that winners and losers with conflicting purposes are sorted out, in such a way that it sustains and strengthens the process itself i.e. the pursuit of interlocked purposes through the market process, permits and stimulates the overall process to continue and enlarges our desires and possibilities. Emery (1977) expresses this in a more general way: '… men will increasingly choose and more consciously strive to choose those purposes that manifest intentions calculated to stimulate both themselves and others to expand their horizons of desire and to rationalize conflict' (p76).

Maladaptive strategies are examples of behaviour that constrains future desires and possibilities whereas adaptive strategies open up possibilities. The principle is a type of extension to Alderson's power principle mentioned above. It is not just the individual actor that seeks to act so as to maximize its own ability to act but to act and interact with others in such a way that enlarges the resources, capabilities and desires of all interacting market actors. Perhaps we should call this the *social power principle*.

**The Extended Firm and the Role of Relationships and Networks**

Lane and Maxfield's (1996) The concept of generative relationships (Lane and Maxfield 1996) and the values and ideals identified by Emery (1977) highlight the importance of relationships and networks in coping with Type 4 turbulence. Further support for the role and importance of relationships and networks comes from the Principle of Requisite Variety (Ashby 1954), where it is argued that in order to respond to its environment, systems have to be able to match its variety or complexity, otherwise they will eventually confront conditions with which they cannot cope. Over time environmental complexity in general has progressively increased, due to the increasing number of people and organisations, which are ever more interconnected and in which the frequency and speed of interaction and response is also increasing. We have moved from simpler hunter-gatherer societies, through early civilizations and the industrial revolution to modern globalized industries and societies (Diamond 1999). Such changes lead to the emergence of different forms of organisation to cope with the increasing complexity (Bar-Yam 1997, 2001).

A critical point is reached when the complexity of the environment exceeds that of an individual person or firm (Bar-Yam 1997, 2001). This means that central direction from a single coordinator, no matter how well supported and informed they may be, cannot cope because of limits to an individual's complexity. Similarly, single firms lack the requisite variety to cope with a highly complex environment. However, networks of organisations, in which knowledge, coordination and direction is distributed and behaviour emerges from the interactions taking place among those involved, have far greater intrinsic variety and therefore greater ability to cope with complexity. The business network itself functions as a kind of collective mind that is more 'intelligent' and able to adapt than the individual firms comprising it, yet it is formed out of those firms and the way they interact and respond to each other. The network mind is implicit and embedded in the patterns of interrelations within and among the actors. For example consider the following description of the Toyota production system:

> 'Toyota's knowledge of how to make cars lies embedded in highly specialized social and organisational relationships that have evolved through decades of common effort. It rests in routines, information flows, ways of making decisions, shared attitudes and expectations, and specialized knowledge that Toyota managers, workers, suppliers and purchasing agents, and others have about different aspects of their business, about each other, and about how they all can work together' (Badaracco 1989).

Another example is the identification of productive technological partnerships from among a set of firms as described by Wolpert (2002). Here the aim is to bring together a set of firms with potentially valuable technological complementarities and facilitate the identification of productive relations within the group. The ability to do that depends on the assortment of firms in the group, from which potential collaborations can be identified. These are not properties of the individual members but a joint property of the group (Leonard and Swap 1999).

*Extending Resource-Advantage Theory to Relations and Networks*

Resource-advantage theory (Hunt and Morgan 1997) may be extended to incorporate the role of relations and networks. To explain this, we start by considering the individual firm and its resources and how they contribute to its performance. We then consider the contribution of connections between firms.

Resource advantage theory argues that a firm's market position is a function of the resources it controls. These include tangible and intangible elements and the competences and skills embedded in the people, teams, relations and networks comprising the firm. These resources are combined in complex ways over time to produce the plans, actions and responses that characterise the firm and the way it responds to its environment, including competitors. In turn these plans, actions and responses determine the firm's market position.

The contribution of a particular resource to the firm's competitive position is not easy to assess for a number of reasons. First, the contribution of a resource will vary over time and across markets depending on the type of value to be created and delivered and the technologies available. Second, the contribution depends on the distribution of the resource, as well as substitutes, among competitors and the ease with which it may be appropriated, duplicated or substituted for. Third, the contribution of a resource depends in part on the other resources of a firm. This is because resources are not used in isolation; they are combined and used in complex ways over time as firms act and respond to their environment in their struggle to survive and prosper. Interaction effects occur among resources that affect their actual and potential contribution to a firm's competitive position. For example, a superior market research department is of limited value if the firm does not have the ability to translate this research into viable products and services. Lastly, a firm's actions and the results obtained have feedback effects on the mix and value of their resources for use in the future. Some are used up or deteriorate; others grow and are enhanced - such as knowledge and learning.

The problem here is analogous to that of determining the effect of different genes on the characteristics of a resulting organism. The organism's genome may be equated to the assortment of resources that a firm has that enable it to behave in particular ways in order to survive and prosper in its environment. The organism or phenotype that results from the expression of genes over time in an environment is equivalent to the organisation and behaviour of a firm that results from the expression of its resources in its environment. This equating of resources to genes is consistent with Nelson and Winter's (1982) use of the term 'routines' as genes i.e. regular and predictable behavioural patterns. These routines 'include characteristics of firms that range from technical routines for producing things, through hiring and firing, ordering new inventory, or stepping up production items in high demand, to policies regarding investment, research and development.' (p14). Such routines reflect the tangible and intangible resources available to a firm, including the competences and skills of firms in using and combining resources. Moreover, the use of resources reinforces or replicates their use to the extent they result in benefits or 'fitness' for the firm as a whole and are replenished by the actions of the firm. For example, the use of a firm's knowledge and research skills to produce competitive marketing strategies reinforces the use of and replenishes these knowledge and research skills.

Resources, like genes, do not act in isolation but interact over time as they are used in combination to create a type of firm with a particular organisation and behaviour repertoire. Firms are selected for based on the fitness of their behaviour in the relevant market environment, there is no absolute standard for judging the fitness contribution of a resource or gene. Fitness depends on the context, including the other resources of the firm and the resources of other firms and organisations. Like genes, individual resources contribute to (or hinder) fitness and survival in the presence of other resources, which in turn contribute (or hinder) fitness in the presence of them. In one context a resource may be extremely valuable in another irrelevant or harmful, such as the value of a market research department in a random placid environment or a perfectly competitive market compared to a clustered or disturbed reactive environment.

A simple example will serve to illustrate how resources are interconnected within and across firms. It is adapted from biological examples described by Dawkins (1983). Resources can be used in various ways by firms depending on their objectives and the other resources at their disposal. Consider how we may determine the effect of a particular resource, A, e.g. a design department, on the competitive market position of a firm. First, we may imagine comparing firms competing in similar markets with or without this resource and see how well they do in terms of market share, price premium, or customer loyalty etc. We may find that resource A is associated with gaining a price premium. Is resource A the resource 'for' price premiums? As Dawkins notes, we can only determine the effect of a proposed cause by comparison with another potential cause. Let us assume there are two firms with resources A1 and A2, which can be thought of as two kinds of design departments. To keep it simple assume there are two kinds of firm outcomes - a price premium or a price discount (P1 and P2). In order to make the comparison we will assume all the other resources of the firms are the same and the two firms are not competing against each other. If we find that, statistically, firms with A1 are more likely to get price premiums (P1 rather than P2) than firms with A2 we may conclude that A1 is a resource leading to price premiums. If price premiums are a key to a superior market position and success in the market, firms with A1 will prosper and other firms may try to build the same type of design department.

Although A1 appears to lead to P1, this may also depend on the other resources of the firm which may suppress or enhance the likelihood of P1. There may be another resource B, the type of advertising department, that interacts with A1. Suppose there are two kinds of advertising departments B1 and B2, and B1 may not get on well with design departments, particularly A1. The presence of B1 may interfere with the activities of A1 directly, or they may be unable to reflect the designs of A1 well in their campaigns. Hence, if all firms have A1 but do not get a price premium if they have B1, then B1 not A1 becomes the resource for gaining a price premium (P1). In other words both resource A and B are potential causes of price premiums depending on the resource assortments that exist in the population of firms.

We may extend this argument to relationships and networks. The contribution of a resource not only depends on the other resources within the firm, it also depends on the resources of other firms with which a firm is interconnected. In this way one firm's resources may find their expression in a related firm's organisation and behaviour, not just their own organisation and behaviour. What we define as the phenotype associated with any genes, or in our case resources, is arbitrary as the chain of cause and effect links extends beyond individual bodies or organisations, to their behaviour and to other bodies and organisations; to what Dawkins (1983) refers to as the *extended phenotype.*

A firm's resources are indirectly combined with the resources of connected other firms and organisations resulting in impacts on the organisation and behaviour of these connected others. Hence a firm's resource may not just affect its own competitive position but also the competitive position of others. This is clearly the case when firms outsource different functions to specialist firms in order to make use of their resources and hence a supplier's resources contribute indirectly to the organisation, behaviour and competitive position of its customers. For example, a supplier's production and design department enhances or hinders the work of the customer's internal production and design department and in turn their relations with other departments such as advertising. The type of relationships a firm has with its suppliers affects the contribution of the supplier's skills and competences to the organisation and behaviour of the customer.

The networks of relationships in which firms are embedded may be viewed as extended phenotypes in which the resources of the firms comprising the network directly and indirectly contribute to its pattern of organisation and behaviour as well as to the reinforcement and reproduction of the resources required to continue this pattern depending on which patterns succeed. To use Dawkins' (1983) phrase this is '*resource action at a distance'* – not within one firm but across firms. Furthermore, the resources of other firms that influence a particular characteristic of the extended phenotype may be in conflict or support each other, just as they do in the case of genes. All the firms involved are not necessarily cooperating with each other; there is still room for conflict in terms of the resource impacts and interlocking objectives of those involved. The extended phenotype is jointly manipulated, not necessarily in a cooperative manner, by the resources, actions and interactions of directly and indirectly related firms who may span different types of industries and technologies.

The important point is that the unit of analysis moves beyond the individual firm to the network of interconnected firms that together co-produce the behaviour and organisation of the network. The survival and performance of individual firms is based on the organization and behaviour of this extended phenotype. Dawkins's (1983, p.233) central theorem of the extended phenotype is that: 'An animal's behaviour tends to maximise the survival of the genes 'for' that behaviour, whether or not those genes happen to be in the body of the particular animal performing it.' Analogously, we postulate that a firm's behaviour tends to reinforce the resources that enable that behaviour, whether or not those resources happen to be part of that firm's resources or some other connected firm's resources.

Natural selection in the market works on the basis of outcomes. The outcomes of firms' plans, actions and responses have feedback effects on their resources, including their skills, competences and knowledge, as well as on their objectives and future plans and interactions. Firms, relations and networks are selected for and evolve based on the outcomes of their actions and interactions, which in turn affects the conglomeration of resources underlying network behaviour. The network, its membership and its resources change and evolve over time. New types of firms with different characteristic interaction and response patterns emerge as well as new types of relations.

Through extended phenotypic effects the interaction behaviour of firms can be indirectly affected by the resources and behaviour of connected firms, which can result in patterns of interactive behaviour becoming instituted in a network that facilitates or inhibits cooperative behaviour. Examples of this are to be found in studies of the evolution of cooperative behaviour among selfish actors in iterated prisoner dilemma games and other types of interaction situations (e.g. Axelrod 1985, 1997; Lindgren 1997; Padgett 1997; Sanchez and Quester 2004). The survival of such cooperative patterns of behaviour depends on whether it is evolutionarily stable, i.e. whether it can survive despite shocks such as 'invasion' by other types of strategies, including those that are more adversarial or non-cooperative.

The extended phenotype view of firms implies a need for a change in management philosophy away from a focus on marketing engineering for the individual firm as an independent market actor and on adversarial relations, to one that presumes the existence of interdependent others and where trust and openness are strongly positively valued (Emery 1977). This appears naïve at first. Surely any firm openly espousing such ideals would be too easily taken advantage of because of the dependence that would be created on others. This is indeed the case because the ability of a firm to prosper through the pursuit of more cooperative strategies depends on the strategies of other firms in the population. So how can such ideals and strategies emerge in a population of firms? There are two basic types of models (Dawkins 1983). Model 1 focuses on the relational or network level. We can envisage a number of competing relationships or networks with different combinations of resources and associated degrees of cooperation and coordination and argue that selection will favour the more cooperative nets in the long run. But such a model does not explain how cooperative strategies came to be effective. In Model 2 the focus is on the individual firm and its resources. Selection in turbulent environments will tend to favour firms with resources, including relationship and networking skills and competences that enable them to interact more harmoniously with other firms in the population, including existing and potential members of their networks. This type of selection mechanism, in which resources are selected for based on the frequency of other resources in the environment, is an example of what is called frequency-dependent selection in population genetics (Dawkins 1983). For example, social insect colonies are extreme examples of cooperatively interacting entities whose genes have been selected for based on their ability to harmoniously and productively interact with other genes in the insect colony's gene pool as well as with external environments. Closer to business, frequency-dependent selection has been demonstrated in the simulations by Axelrod (1980, 1997) and others that show how groups of cooperators can emerge and prosper over time in populations of selfish competitors.

Relationships involve a joint choice and investment; firms need to be chosen as well as to choose partners, unless they are in very powerful positions. Hence part of the strategic problem, for a firm is to compete to cooperate, to be able to join and remain in viable relations and networks (Wilkinson and Young 2002). In order to do this it has to devote resources to attracting relationship partners, becoming attractive as a relationship partner and heading off rivals - the business equivalent of sexual selection in biology (Wilkinson et al 2003). As Darwin (1874) noted in his second great work, much of biological evolution is unexplainable simply in terms of natural selection – the ability to survive and adapt to local conditions. There is also a struggle to reproduce in order for genes to survive into the future generations and this depends on finding sexual mates and ensuring the survival of your own progeny. Features that do not appear to be advantageous in the business of living, and in fact may be downright harmful (e.g. long tails), may evolve to increase attractiveness and hence aid sexual selection. In the same way the evolution of viable relations and networks in turbulent business environments requires the evolution of firms who can establish cooperative partnerships, which may make them both more vulnerable as well as more attractive.

The rise of relationship marketing and concepts of collaborative advantage, partnering and the like suggests that the population of firms and their resources is changing in favour of those with more cooperative abilities and more cooperative firms will survive in populations of cooperators. However, from an individual firm's perspective, adopting cooperative strategies is still not easy, especially if they begin from a history of previously adversarial relations. But there are studies reporting how firms have attempted to change the nature of their relations with key others, starting with small, less risky actions to build a cooperative trusting relation (e.g. Corbett, Blackburn and van Wassenhove 1999; Holmen, Hakansson and Petersson 2003; Welch et al 1998). As these cases make plain, relationship building takes time and is an area requiring more research attention.

*Soft Assembled Strategies*

The foregoing shows that the complexity of turbulent environments requires firms to reach out beyond their boundaries and build relationships to access, build on, and to evolve understanding and response capabilities. Survival depends on the adaptation of the overall relational network, including changes in the nature of individual actors and their links. There is no guarantee that a particular focal firm will survive as a recognizable entity in the future, but through its actions and responses it will have played a part in the evolutionary and self organizing process taking place; similar to the way biological evolution occurs. The problem for the firm is therefore to participate in the game of business through its interactions and responses to others in such a way that the game evolves in ways that it continues to be included as a viable player. But as the nature of evolution of business ecosystems cannot be foreseen; there are no magic strategies or prescriptions that can guarantee future survival.

This is not a very comfortable type of normative theory of management because it gives no clear guides for action and results in a number of paradoxes (Hakansson and Ford 2002). Under these circumstances the role of management and strategy in firms changes from one of designing optimal relationships and strategies to what may be described as *soft assembled strategies* (Andy Clark 1997). Although developed in the context of individual behaviour and cognition, we believe this concept has direct counterparts in business and marketing. The extended view of the firm is similar to the concept of the embodied mind in cognitive science Clark (1997), in which the mind, body and local environment are viewed as part of a dispersed and extended mind that is the sensing, thinking and responding system of a person. Thus, we sense and think in part with our body and with parts of our local environment. We do not, for example, solve the problem of picking up a piece of paper by issuing commands to our muscles to act; we enlist the innate characteristics of our arms and fingers to solve the problem. The technologies, tools and equipment we use, including language and other people, become part of our extended self in the same way that the stick of a blind person becomes part of them and the way they sense and respond to the world. They do not simply extend and enhance existing skills and abilities or remedy a defect but provide new forms of functionality, creating new niches for action and intervention. For example, e-mail is not a replacement for face to face communication: 'It provides a *complementary functionality*, allowing people informally and rapidly to interact, while preserving and inspectable and revistable space.' (Clark 2003 p110, emphasis in original). Clark argues that humans are natural-born cyborgs able to expand and reinvent our sense of body and action, that our body image and boundary is highly negotiable, and our brain's plasticity enables it to learn to exploit new kinds of feedback loops and action potential that come from being linked to external systems and technologies. We sense, think and act <u>with</u> our environment by learning from it and drawing on its intrinsic response tendencies such that our body and the local environment become part of the means of solving problems not their source. As Clark (1997) summarises it:

> 'The job of the [central nervous system] … is *not* to bring the body increasingly 'into line' so that it can carry out detailed internally represented commands directly specifying e.g. arm trajectories. Rather the job is to learn to modulate parameters (such as stiffness), which will then *interact* with intrinsic bodily and environmental constraints so as to yield desired outcomes. In sum, the task is to learn how to soft-assemble adaptive behaviors in ways that respond to local context and exploit intrinsic dynamics. Mind, body, and world thus emerge as equal partners in the construction of robust, flexible behaviors' (p45, emphases in original)

In much the same way we can view the manager and individual firm expanding and reinventing their sensing, thinking and responding potential with and through their internal and external networks of relationships. To adapt the words of Clark (1997) to business we argue that:

> 'The job of the manager is *not* to bring the firm increasingly into line, so that it can carry out detailed plans centrally developed, such as how to position and carry out marketing activities for a product or service. Rather the manager's job is to learn to modulate parameters of the organization, such as its customer responsiveness, its relationship and network competences, which will then *interact* with intrinsic firm, relationship and network constraints so as to yield desired outcomes. In sum, *the task is to learn how to soft assemble adaptive behaviours* in ways that respond to local intra and interfirm context and exploit intrinsic dynamics. Managers, firm and network thus emerge as equal partners in the construction of robust, flexible behaviours.'

The example of ROLM referred to above and the way it identified and participated in various types of generative relations as it felt its way forward in the PBX market is an example of a soft assembled strategy. The original strategies and interpretations of the market of ROLM and others changed in response to participation in relations and, at the same time, changed the pattern of relations. Soft assembled strategies are also reflected in the ways lead users have been incorporated into the innovation process (Franke and von Hippel 2002; von Hippel, 2001), such as the development of chef packs to enable meals designed by chefs to be easily mass-produced (von Hippel, 2001). The problem was that of scaling up production of new types of meals designed by chefs in their kitchens for catering firms, airlines and packaged gourmet meals. Chefs had worked with ingredients that were not identical to the massed produced ingredients used in the manufacturing process and so it became a major problem to reproduce what the chef did in their kitchen. The solution was to develop 'chef packs' of the bulk ingredients, so that when chefs invented new meals using these ingredients – and they do this by tasting and experiment – the meals were immediately mass producible. Similar types of solutions are to be found in the way the Apache public domain web site developer is developed by users and then incorporated into advanced versions of the software and in the way the Unix operating system Linux is developed.

Iansiti and Levien (2004) describe the way firms play different roles in the business ecosystems involved in developing computer hardware and software systems. One type of role is that of 'keystones' who are richly connected hubs providing the foundation for creating niches and increase diversity and productivity in their business ecosystems. TSMC is a keystone in the integrated circuit ecosystem and NIVDIA is a successful niche player. NVIDIA outsources all fabrication of its graphics processing units to TSMC, thereby leveraging the expertise of these suppliers. TSMC is based in Taiwan but enables its clients, like NVIDIA, to maintain close contact with the production system for its chips, to the extent that the client can make late engineering changes and even cancel orders at the last minute without a large penalty. TSMC uses the Internet to make information on designs and products available from its technology library all the time so that the customer can access what they need without human intervention. They do not do any designs themselves or compete with their customers in design; its systems are set up to let its leading edge customers lead. It would be impossible and impractical for the enabler to try to fully understand its customers' requirements and offer them potential solutions. Instead, it provides them with the necessary tools and means *for others to use it* effectively in solving their problems as they arise.

In these examples firms attempt to develop relationships with customers and other firms that are generative, in which they can play both leadership and followship roles; allowing themselves to be planned by others as well as to plan for others, to change their interpretations of their environment and to harness their resources in ways they could not anticipate. It is impossible to capture the customer's or other organisation's understanding or 'mind' even with sophisticated marketing research and modelling methods, and to transfer this understanding to the 'mind' of the supplier.

Learning to soft assemble strategies limits the need for managers to explicitly try to take into account all the direct and indirect relationship and network effects of their resources and actions. This is impossible and it is likely to be counterproductive even if it were possible because it would make resources and networks even more richly interconnected and more unpredictable and chaotic. For this reason Kauffman (1994) has suggested that evolution has tended to produce organisms that are *optimally myopic strategists:* 'not because of costs of computation, but because, if we are individually too clever, we tend to transform the world in which we are adapting into a yet-more-chaotic world in which we fare less well.' (p 84)

For firms, soft assembled strategies involve two basic components: (a) positioning and repositioning a firm in networks of relations (Johanson and Mattsson 1992) and (b) exploiting and exploring relationship and network opportunities, threats and action potentials (March 1991). The former focuses on the way a firm couples itself to other firms and organisations both directly and indirectly and thereby extends the mind and resources of the network i.e. the sensing, thinking and action potentials of the network, or variety in Ashby's (1952) sense. Positioning involves micro positioning in individual relations and macro positioning in networks of relations. The coupling process is not under the direct control of the firm because it involves a double choice, choosing and being chosen by others, and acting and responding. In addition it is as much about enabling others in the network as the focal firm, because we cannot know in advance where new opportunities and threats will arise, who can and will recognise them and how to respond. There is no ideal network of relations. At any time a firm is involved in a portfolio of many types of relations with varying functions and connections to other relations, mixtures of cooperation and competition and different strengths, which have evolved over time through the actions and interactions of those involved. Both strong and weak ties are important. Weak ties, because they are associated with structural holes (Burt 1992), are potential sources of new types of information and play a key role in the discovery of new opportunities (Kirzner 1997). Strong ties are important because they are the means by which opportunities or threats can discover a firm (Yanto et al 2005) as well other functions already discussed. Strong cooperative ties are the means by which information about such opportunities and threats are passed on to others or shared and collaborative action and response is orchestrated. Exploitation and exploration strategies involve trading off the gains to be made from working within existing relations and networks versus looking for and developing new potential relations and networks. Existing relations and networks both enable and constrain what can be sensed, understood and responded to by the network and these develop and evolve over time through the actions, interactions and outcomes taking place. The way a network responds to particular challenge depends not only on the nature of the challenge but on its history and pattern of relations, as these set the stage, the starting conditions, for encountering and exploring the future. From a society's perspective a variety of networks is needed in order to ensure that it is not locked into only one way of seeing and responding to the future. The individual firm and its people have to be sufficiently flexible to continually re-sculpture their network role and position to remain viable players in viable networks and, in so doing, contribute to the evolvability of the networks in which they participate.

**Conclusions and Research Implications**

We have shown how relationship marketing and network strategies play different roles according to the type of environment in which they operate and that this leads to different types of normative theories for management in terms of how to identify, produce and sustain productive interfirm relations and networks. To date, most attention in relationship marketing has been focused on its role in Types 1, 2 and 3, in which the firm is the primary strategic actor and relations with customers and other organisations can be valuable means by which firms create and sustain competitive advantage. But the types of business problems facing firms in type 4 environments are beyond the capacity of individual managers and firms to sense, understand and respond to.

The increasing number, degree and speed of interconnectivity among people and organisations has produced environments that are complex and turbulent, in which firms face strong forms of uncertainty. The future is unknown and unknowable because it is produced by complex patterns of interactions over time that make it highly non-linear and beyond analytic solution. To survive and adapt to such an environment requires a firm to effectively participate in, help co-create and sustain networks of relations in which no firm has control. This requires a change in management ideals, values and strategies. The forces that have produced this richly inter-connected world provide some of the means for people and firms to renegotiate their boundaries and bring the manager, firm and network closer together. This includes developments in information and communication technologies that enrich network, relationship and interrelating possibilities. Such developments not only extend and enhance existing firm resources and functionalities, they create new types of complementary functionalities, as managers learn to exploit different kinds of feedback loops and action potentials that come from being coupled to other firms and organisations in new ways. The network extends more than the body of the participant firms, i.e. their resources and action potential, it extends their mind and senses so that, through interactions with others, firms are able to recognise, comprehend and respond to opportunities and threats they otherwise could not.

In addition, there are signs that changes in business values and ideals are taking place, as firms are becoming more aware of and sensitive to the wider impacts of their actions and are increasingly exploring the role of more collaborative strategies, by bringing suppliers, customers and other organisations earlier into strategy planning and development processes. But including other firms thinking in your own planning and strategy development is only part of the problem. They also need to be included in the planning and strategy development processes of relevant others. To achieve this requires an ability to adapt both goals and means (not just the means for achieving fixed goals) in response to others, to be an effective follower as well as leader and to be prepared to be 'planned' by others as well as to plan for others.

Strategy becomes more about searching via interaction processes to discover and respond to opportunities and threats, rather than about designing and controlling resources to achieve desired outcomes. Network positioning and repositioning become key strategic issues as firms seeks to identify and engage in generative relations through which they sense, interpret and reinterpret their world and act and respond to it (Gadde, Huemer and Hakansson 2003; Hakansson, Harrison and Waluszewski 2004). The relevant networks and relations here are not confined to a particular industry, place, or nation, though these may be important, instead they may span industries, technologies and nations (Wilkinson, Mattsson and Easton 2000).

Methods developed for guiding and solving management problems in other types of environments are of limited value in dealing with Type 4 problems. Marketing engineering approaches may be able to solve part of the problem but, in turbulent environments, such optimisation problems are embedded in more complex problems that have no optimal solution and that relate to the broader environmental and strategic context in which the firm operates and in which the strategy will be understood and implemented (e.g. Lilien et al 2004). Local optima may be poor solutions for both the individual and their relationships in the wider context and may serve to limit a firms understanding and ability to respond.

*Research Implications and Agent Based Models*

The focus of this paper has been on the strategic problems confronting firms in different types of environments, particularly turbulent environments. In order for us to improve our understanding of this type of environment, to sensitise managers to the complexities involved and to give them better advice about identifying and responding to such environments further research is needed. There is a fast growing area of complexity science that focuses on the role and impact of interactions in shaping the dynamics and evolution of complex adaptive systems, such as type 4 environments. This offers both theories and methods to advance our understanding and responses to such environment in marketing contexts. While there are signs of a growing interest this area in marketing is still underdeveloped compared to other disciplines (e.g. Holbrook 2003). Here we offer some suggestions as to how marketing can make better use of its theories and methods.

The prime method employed in complexity research is agent based simulation models to create artificial worlds mimicking key features of a focal system (e.g. Langton 1996; Casti 1997). Such models are not based on a set of general driving equations, as was the case in previous types of simulations (e.g. Forrester 1966), but allow for a heterogeneous interacting set of agents, which could represent firms, individuals or other objects, with their own goals, resources and rules of behaviour that may change as a result of learning. Relations and networks are created through ongoing interactions. No equilibrium outcomes are assumed, instead the system gravitates towards different types 'attractors' depending on various model parameters.

Studies of actual business systems reveal only one history, what types of relations and networks have been successful and how they developed. But this does not tell us whether this is one of a number of possible histories and types of outcomes or not. Management needs to know which factors influence the likelihood of different types of outcomes emerging and the role and impact management interventions can have. Such an analysis is impossible when we only have one or a few detailed histories to use; we cannot go back and run history again to see how sensitive it is to different factors. But agent based simulation models enable us to capture important features of the evolutionary process and conduct experiments to see identify alternative types of outcomes, tipping points, and how sensitive and stable they are to different conditions and disturbances including various types of management interventions.

McKelvey (2004) sees agent based models as offering a middle ground between thick and thin descriptions that is capable of bridging the two. Thick descriptions result from in depth case studies of actual business histories, which reveal some of the complex causal processes involved but which cannot be easily generalised. Thin descriptions result from sample survey type research which is more generalizable but which abstracts away from any examination of the processes, events or choices by which different types of variables are interrelated and affect outcomes (Abbott 1992).

> '[R]esearchers (1) could begin with agent models that generate multifinality samples, (2) find abstracted solutions cutting across the multifinality variances, and then 3) model these abstracted, potentially equifinality solutions under different environmental conditions to test how much generality they have and more specifically, identify conditions where they do or do not hold true.' (McKelvey, 2004, p. 333)

This is the challenge for marketing theory and research if we are to make headway in developing normative marketing theory for managers in complex adaptive systems. Agent -based models offer the opportunity to develop and test our normative theories, sensitize managers to the perplexing behaviour of non-linear complex systems and to advance our theories of the dynamics and evolution of marketing systems and practices.